# The Role of Multiferroics in the Negative Index of Refraction

David W. Ward[1], Eric Statz[1], Kevin J. Webb[2], and Keith A. Nelson[1*]

[1]*Department of Chemistry, Massachusetts Institute of Technology, Cambridge, MA 02139*
[2]*School of Electrical and Computer Engineering, Purdue University, West Lafayette, IN 47907*

(Dated: January 1, 2004)

## Abstract

We explore the possibility of realizing intrinsic far infrared negative index materials (NIM) in multiferroic crystals (crystals simultaneously possessing a ferroelectric and ferromagnetic phase) possessing electric and magnetic dipole resonances with nearby resonance frequencies, or overlapping regions of negative permittivity and permeability. We demonstrate the functionality of such a material using finite difference time domain simulations. In order to motivate the connection between multiferroics and negative index materials, we discuss the negative index of refraction in the polariton picture.

## Introduction

Materials simultaneously displaying negative permittivity and permeability are presently a hot topic of debate. These materials are often referred to as negative index materials (NIM) because it has been shown that the interaction of such materials with electromagnetic radiation can be described by a negative index of refraction[1]. Some of the most striking consequences of this are a reversal of the phase velocity with respect to the group velocity, reverse Doppler shift, reverse refraction, and sub-diffraction limit imaging, i.e. a perfect lens. To date, experimental realization of NIM's has only occurred in metamaterials composed of high frequency electrical and magnetic resonant reactive circuits that interact in the microwave band[2]. We aim to extend NIM studies to the far infrared, and below we describe how it is that we model far infrared NIM and illustrate the importance of the mechanical phase components in the wave impedance which controls the phase between E and H and dictates power flow in a material. Further, we propose that multiferroic crystals may have potential application as a NIM.

## Results and Discussion

Much attention has been given to the phase velocity, $v_p = (\varepsilon\mu)^{-1/2} = c_0(|\varepsilon||\mu|)^{-1/2} e^{-i(\phi_\varepsilon + \phi_\mu)/2}$ in a negative index material, where we have introduced the magnitude and phase of the permittivity and permeability. It is clear then, that a negative phase velocity is experienced when the sum of the phases from the magnetic and electric response is $2\pi$. Less attention has been given to the wave impedance, $Z = \sqrt{\mu/\varepsilon} = (|\mu|/|\varepsilon|)^{1/2} e^{i(\phi_\mu - \phi_\varepsilon)/2}$ which determines the phase between the electric and magnetic fields and hence the direction of energy propagation and degree of attenuation. For example, consider one-dimensional propagation of an electromagnetic wave from vacuum into a material near, but below, an electric dipole bandgap (or Restrahlen region). In this case, the mechanical component in the Poynting theorem, coupled to the electromagnetic field through the

---

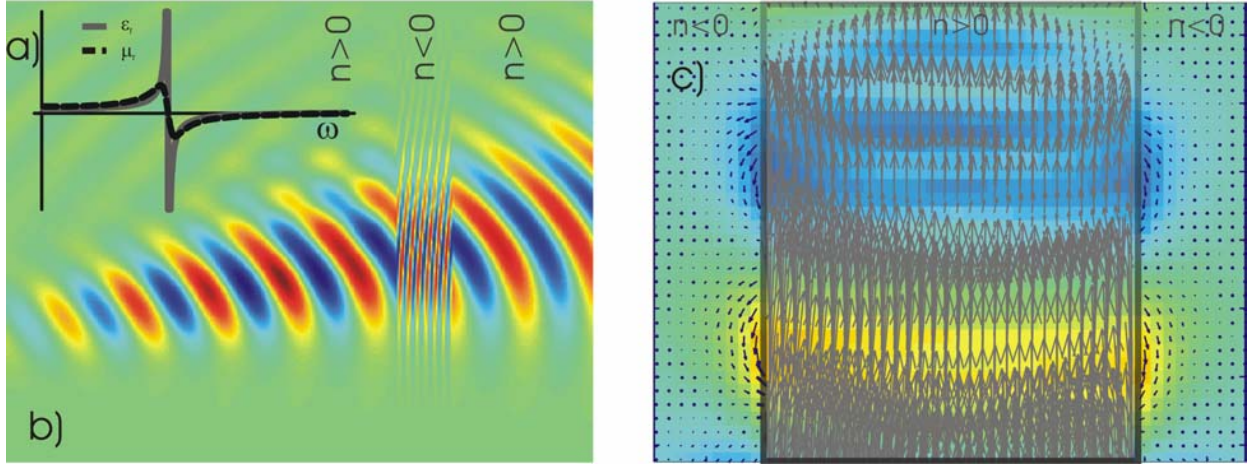

**Figure 1:** a) Graph of permittivity and permeability that meet the NIM criteria. b) Negative refraction at the interface of a positive index and negative index material. c) Reverse power flow near the boundary of a waveguide as imposed by the material boundary conditions.

electric dipole coupling, is almost $\pi/2$ out of phase with the field, and if the permeability is simply that of free space, then E and H are nearly $\pi/2$ out of phase resulting in strong damping. Above the bandgap, at $\pi/2$ for example, the Poynting vector goes to zero or is opposite to the incident radiation resulting in exponential decay instead of propagation. Now, imagine that we introduce a magnetic mechanical component that interacts with the impinging electromagnetic field through a magnetic dipole coupling. If the magnetic bandgap is the same as the electric bandgap then the H field is retarded compensating for the retardation in E introduced by the electric bandgap. Now, instead of total reflection at the interface we have propagation, but with the phase velocity opposite that of the group velocity since the Poynting vector is restored to the forward direction. We can imagine extending this argument to two or three dimensions and describing refraction at an interface in terms of the phase of the H or E components for TM or TE radiation and the concomitant change in the direction of propagation introduced by the phase of the E and H components.

Since the retardation of E and H is conveniently described in terms of the phase of the permittivity and permeability, it is illustrative to think of the phase of the mechanical component instead, which is analogous to the response of damped oscillators. The interaction of electromagnetic radiation with a spatial array of independent mechanical oscillators coupled to the electromagnetic field through an electric or magnetic dipole interaction is referred to as the polariton picture[3]. The elementary excitation, polariton, in the vicinity of the mechanical resonance behaves much like the surface of a water bed—the temporal wave motion is linked directly to the spatial amplitude distribution. Natural sources for these polariton resonances are found in ferroelectric and ferromagnetic crystals. In this case, the polaritons are referred to as phonon-polaritons and magnon-polaritons, respectively. This suggests a natural source of negative index materials may likely be found in multiferroic crystals with the electric and magnetic resonances in the same spectral vicinity. Magnetic effects can be substantial at low frequency, but tend not to be at optical frequencies; however, ferromagnetic and antiferromagnetic materials are known that have resonances as high as a few THz. It has been proposed that the scarcity of multiferroics in nature is due to competing effects surrounding *d* shell electrons in the crystal, but a few natural sources are known and fabrication of new multiferroics is underway[4].

Due to the lack of real model multiferroics, we model with a hypothetical one. Specifically, we consider the situation in which both a ferromagnetic and ferroelectric resonance is present (multiferroic) with overlapping negative permittivity and permeability (negative index criteria) (fig. 1a). Using this model, we simulate negative index materials in the THz region of the electromagnetic spectrum and demonstrate some interesting effects that could be experimentally demonstrated if the proper multiferroic is found or fabricated. Specifically we illustrate refraction (fig. 1b); signal, front, group, phase, and energy velocity; power flow in waveguides (fig. 1c); and material dispersion relations in a hypothetical multiferroic negative index material using FDTD in the polariton picture[5].

## Conclusion

In conclusion, it is the ability of the mechanical phase in a ferroelectric or ferromagnetic material to control the phase between E and H and hence the direction of power propagation that identifies multiferroics as a possible source for a negative index of refraction, presenting new opportunities for the multiferroic community.

## Acknowledgements

*Funding was provided in part by the MRSEC program of the NSF under grant no. DMR-0213282. The authors thank John D. Joannopoulos and Chiyan Luo for helpful discussions.*